\documentclass[12pt]{article}
\usepackage[dvips]{color}
\usepackage{graphicx}
\usepackage{amsthm,amssymb,amsmath,graphics,times}
 \usepackage{amscd}
\usepackage{color}
\usepackage{url}
\usepackage{CJKutf8}
\usepackage{fullpage}
\usepackage{comment}
\usepackage{url}

\usepackage{hyperref}

\newcommand{\mean}{{\rm mean}}

\newcommand{\foredays}{n_{\rm f}}
\newcommand{\matcdays}{n_{\rm m}}

\newcommand{\weightp}{\tau_{(k-z)}(+z)}
\newcommand{\weightsuc}{\tau^{suc}_{(k-z)}(+z)}

\begin{document}







\centerline{\large \sffamily Forecast U.S. Covid-19 Numbers by Open SIR Model with Testing}


\medskip
\centerline{\sffamily Bo Deng\footnote{Department of Mathematics,
University of Nebraska-Lincoln, Lincoln, NE 68588. Email: {\tt
bdeng@math.unl.edu}}}

\noindent
\textbf{Abstract:
The U.S. Covid-19 data exhibit a high-frequency
oscillation along a low-frequency  wave for outbreaks.
There is no model to account for it. A modified SIR
model is proposed to explain this spiking phenomenon.
It is also used to best-fit the data and to make
forecast. For the simulated duration of 590 days, the model
is capable of achieving a 0.5\% mean squared relative
error (MSRE) fit to the seven-day average of the
daily case number. The outright 28-day's prediction by the model
generates a 20\% MSRE for the cumulative case total
due to a persistent underestimation of the data by the model.
With the proposed correction to the abberation, the model is
able to keep the 28-day's cumulative case total forecast
within 10\% MSRE of the data.}

\bigskip

The standard epidemiological model (\cite{Brau2013,Vand2002})
for the spread of an infectious disease
in a fixed population $N$ is the SIR model:
\[
S'=-cSI,\ I'=cSI-rI,\  R'=rI, \  S+I+R=N
\]
where $S$ is the number of susceptible at time $t$, $I$ the infected,
$R$ the recovered, and $S'=S'(t)$ is the rate of change
(derivative) of variable $S(t)$. Parameter $c$ is the per-infected infection rate
and $r$ is the recovery rate. From the onset of Covid-19 pandemic,
researchers everywhere shifted into overdrive in search of more realistic models
in order to explain the live data and to predict future trend of the pandemic.
All models fail by one crucial test. Just a few weeks into the pandemic,
it became apparent that the US data exhibit a seven-day oscillation.
At first one could dismiss it as an artifact because there are seven days
in a week. But after the appearance of the Omicron variant in the summer
of 2021, the seven-day oscillation changed to a 3-day oscillation,
and the seven-day-a-week explanation dissolved.
None of the models I saw in the literature is capable of such oscillations
(e.g. \cite{Anas2020,Bert2020,Moei2021,Petr2022,Rahi2020,Shin2020,Zou2020}).
Pandemic forecasting is considered a `wicked' problem by \cite{Luo2021}
and declared a failure by \cite{Ioan2020}.
The purpose of this article is to present a model and to show how we can fit it
to data and then to make reliable forecast on case number and death number.

\begin{figure}[t]

\centerline{
\parbox[l]{3in}{
\centerline
{\scalebox{.5}{\includegraphics{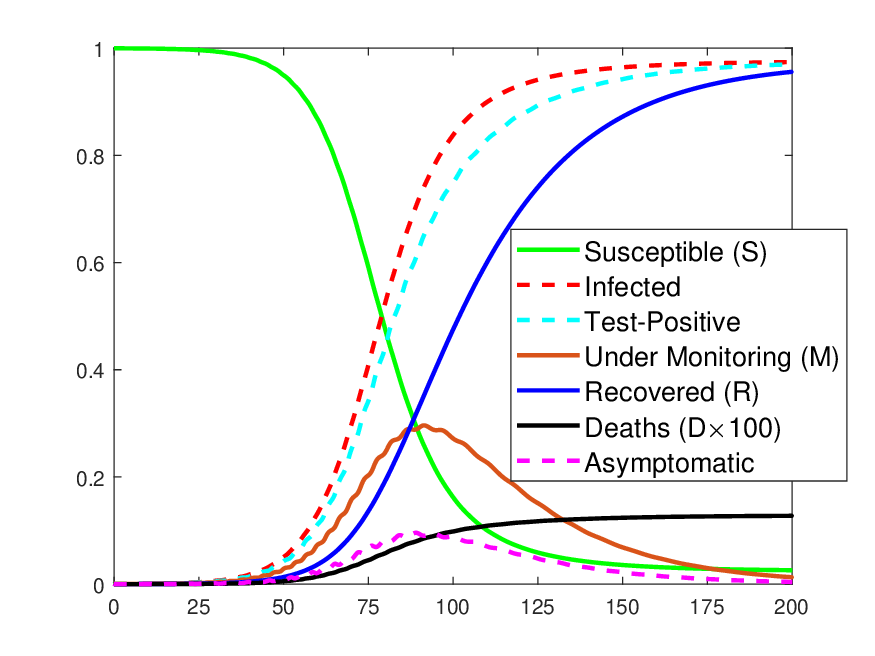}}}
}
\hskip 1cm
\parbox[l]{3in}{
\centerline
{\scalebox{.5}{\includegraphics{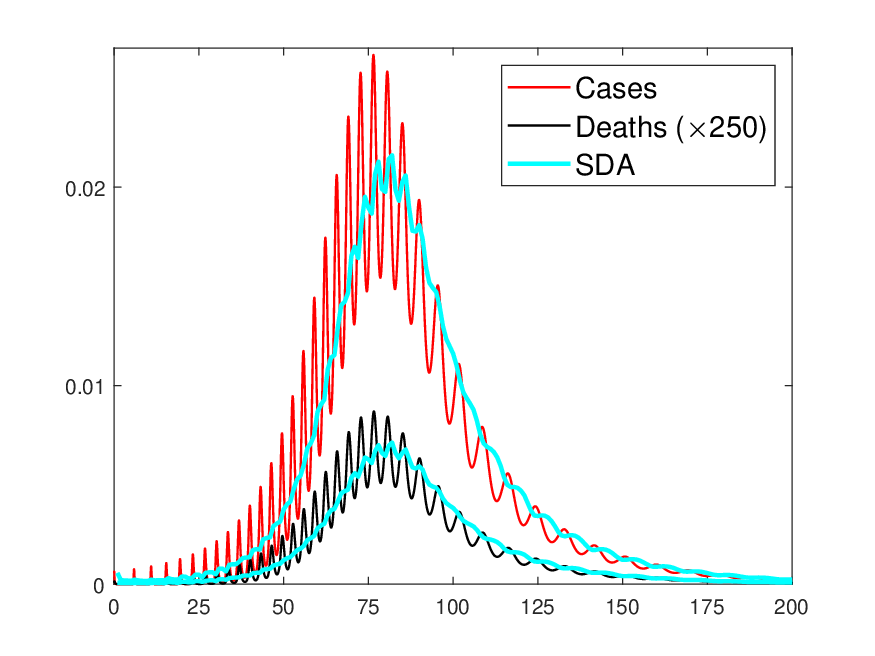}}}
}
}
\centerline{\ \ \ \ (a)\hskip 3.3in (b)}


\caption{\textbf{Model Dynamics:} (a) An example of dynamics for
the SICM model (\ref{SICM}), with parameter values:
$c=0.6,\ p=60,\ a=2.2,\  m=7.6,\ d=0.01,\ q=0.05$, and normalized
initial values: $S=0.99977,\ I=0.0001,\ C=0.00003,\ M=0.0001$.
The `infected' and `test-positive' curves are the cumulative totals
for $I$ and $C$ respectively. Their difference is the
daily number of asymptomatic. The total deaths is plotted
100 times the actual value $D$ to boost its visibility. (b) Daily
numbers ($P$ and $dC$) for the same parameters and initials, with
250 times the daily death number to boost its visibility.
Each has its own seven-day average (SDA) curve.
All variables are normalized against a total
population $S(0)+I(0)+C(0)+M(0)+R(0)+D(0)=N$. The test
rate parameter $p=60$ per day per case translates to 24 minutes
for test time per case. This sample
dynamics is not used to fit any real data. }
\label{figModelDynamics}
\end{figure}

\bigskip\noindent
\textbf{The SICM Model.} There are too many models to list when it comes to
Covid-19 modeling. To the SIR model, people can and always do to include
many compartments, e.g. exposed, hospitalized, deceased, asymptomatic,
young and old, with and without underlying health conditions, etc. Rather than
sorting out the pros and cons of such compartments
we introduce ours below with minimal comments and leave
its justification to how well it can fit and explain the data.
\begin{equation}\label{SICM}
\begin{array}{l}
S'=-cSI\\
I'=cSI-\frac{pC}{I+aM}I\\
C'=\frac{pC}{I+aM}I-mC-dC\\
M'=mC-qM\\
R'=qM\\
D'=dC\\
S+I+C+M+R+D=N_k\le N
\end{array}
\ \
\begin{array}{l}
\hbox{ for data fitting:} \\
\quad (k-\matcdays)\le t\le k \\
\hbox{ for forecasting:} \\
\quad {k}\le t\le (k+\foredays) \\
\end{array}
\end{equation}
Here, in addition to the susceptible class $S$, the infected class $I$,
and the recovered class $R$, we include three more classes: $C$ is
for confirmed cases by testing, from which $dC$ many per-day
go to the deceased class $D$, and $mC$ many per-day go to
the monitored class $M$ for which at least one more test is done
in some future days. At a rate of $q$,
the monitored comes out from monitoring without further testing and
goes into the recovered class $R$. Although we call it the recovered,
it is not the conventionally defined recovered class for the basic
SIR model. It is rather a subclass but a substantial one.
Our $R$ class does not include for example
asymptomatic individuals, nor any individual who is tested positive
but does not go into monitoring by further testing because of mild symptoms.
Notice that a patient who dies from the infection does
not receive another test for Covid-19 unlike those in $M$.
All assumptions are probabilistic. For example, the last
assumption can be stated accurately as that with probability
negligible an infected person who died from the infection did not receive
another Covid-19 test after their initial diagnosis by testing.
Fig.\ref{figModelDynamics} gives an illustration of the spiking
dynamics of the model (\ref{SICM}).

The first date when the Covid-19 case and death numbers was reported
for U.S. from CDC's data is Jan. 22, 2020. The end date for this
study is Sept. 1, 2021 (\cite{CDC21}). There is a total of 590 days in between over which
our model is fitted to the case and death data. Integer $k$ is
between 1 and 590, and is used to denote the $k$th day when forecasting
is simulated, also referred to as `the forecasting day',
`the working day', the `0-day', or `today' throughout.
For the SICM model in the continuous time $t$, we will set the
time unit in day so that $t_k=k$ automatically.

\begin{figure}[t]
\centerline
{\scalebox{.6}{\includegraphics{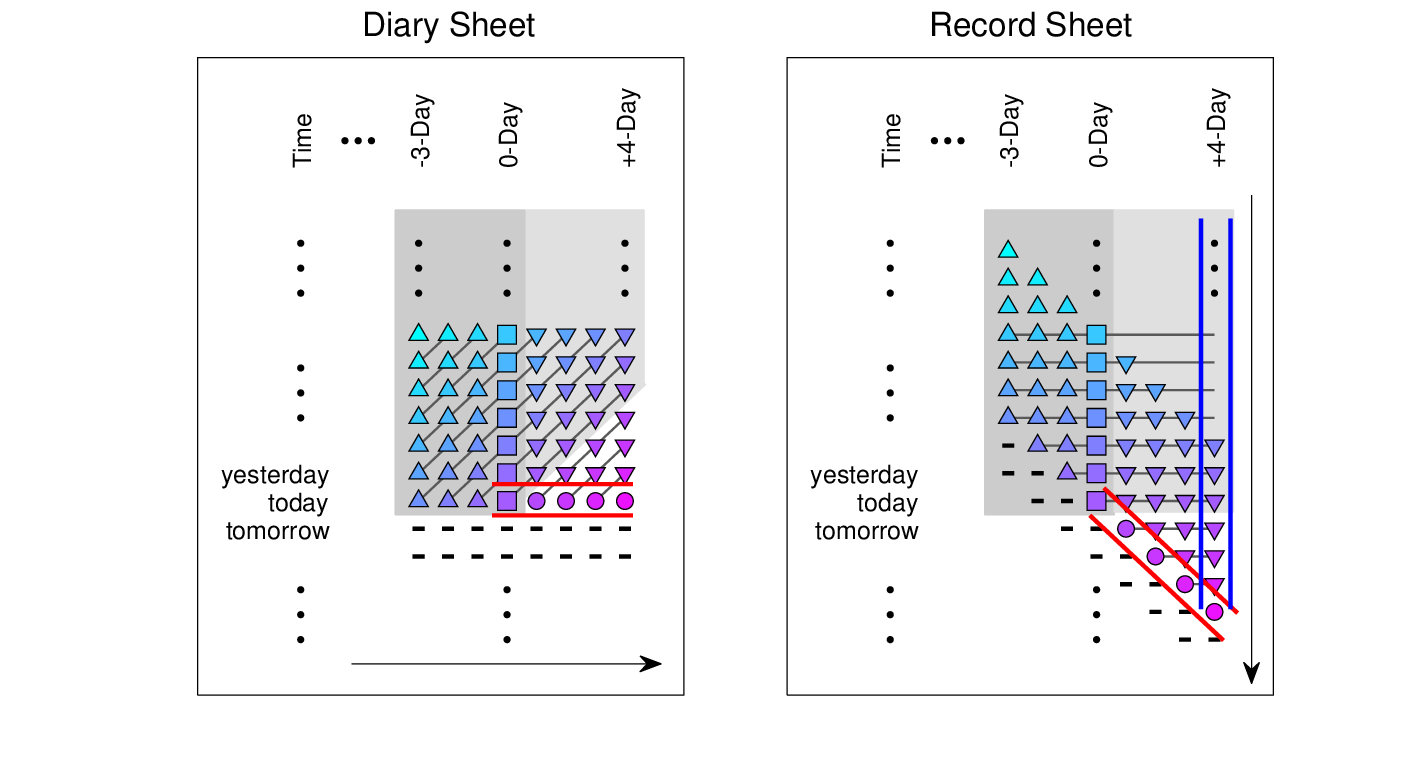}}}

\vskip -.3in
\caption{\textbf{Book Keeping on Forecasting Day:}
Triangles are
for matching, and squares on the 0-Day column
are for matching and real data collection
on each working day. Upside-down triangles
are forecasted numbers,
and discs are forecasted numbers done
on the working day (`today'), which then change
to upside-down triangles when `today'
becomes `yesterday' (or $(-1)$-Day). Entries with a
dash are to be filled on future working days.
For parameters, each row on the Diary Sheet has one
best-fitted value for all `$(-)$' and `$(+)$' days.
Each gray line connects all data entries which
are either fitted values or forecasted values
to a given day's real data represented
by a square on the 0-Day column. For a perfect
model, all values connected by a line are identical.
The record sheet is a rearrangement of the diary
sheet when all values on the slanted lines are
rotated clockwise to become horizontal, for which
the transformation is referred to as realignment.
}\label{figForecastBook}
\end{figure}

Our first assumption for the model is that the parameters are not
fixed throughout the entire 590-day period because of
a variety of reasons: the arises of new variants, behaviour changes
as the result of information feedback, changes in mitigation measures, etc.
Instead, we assume for each forecasting day, the parameters remain constant
only for a duration of $\matcdays$ many days prior to the 0-day. The
prior days are referred to as $(-1)$-day for `yesterday', $(-2)$-day
for `the day before yesterday', etc. We also assume for each $0$-day, the
parameters remain constant for a duration of $\foredays$ many
future days, with $(+1)$-day for `tomorrow' and $(+2)$-day for `the day
after tomorrow', etc. We use $\matcdays=21$ days and
$\foredays=28$ days for the study. Thus, for the $k$th-day's data,
there are 22 working days when it is used to fit the model. These
are the $(k+0)$th working day for its 0-day's fit,
the $(k+1)$th working day for its $(-1)$-day's fit, and up
to the $(k+\matcdays)$th working day for its $(-\matcdays)$-day's fit.
In contrast, there are 28 working days when the $k$th-day's data
is a forecasted value. These are the $(k-1)$th working day for
its $(+1)$-day's forecast, the $(k-2)$th working day for its
$(+2)$-day's forecast, and back to the $(k-\foredays)$th working
day for its $(+\foredays)$-day's forecast.
Fig.\ref{figForecastBook} gives an illustration on how to
organize all data for the study. By definition, a \textit{perfect}
model for a data has the property that for the $k$th-day's data,
all its minus-day's predicted values and all its plus-day's
fitted values are equal. For every imperfect model, these
values are only estimates or approximations of the data.
For the illustration of Fig.\ref{figForecastBook}
each row of the diary sheet lists things done on a working day,
and for the record sheet each row lists things
(fitting or forecasting) done to a given day's data.

The second, and most essential, assumption for our model is that the
total population at each given forecasting day
is not fixed as the entire U.S. population.
This assumption implies that, for example, when the virus first hit New York
City, it did not make people from Alaska susceptible instantaneously
simply because Alaska is a state of U.S.. In another scenario, if one
hunkers down or completely isolates themselves from contacting the virus,
then he or she is not part of susceptible, or at best is only
a fraction of one susceptible. That is, for each fitting and
forecasting period for  $(k-\matcdays)\le t\le (k+\foredays)$,
the rolling total population is $N_k$, a parameter to be fitted,
no greater than the true U.S. population $N$.
Because of this assumption, we call our model
an \textit{open} model in contrast to the basic SIR for which $S+I+R=N$,
i.e. the sum of $S,I,R$ is closed by the total population.
Because the rolling total $N_k$ is treated
as a parameter for each 0-day, subject to model-to-data fitting like
other parameters, the equations for the variables $R$ and $D$ are
decoupled from the rest. That is, to fit the model to the data, we
only need to compute the first four equations in $S, I, C, M$
in order to find the parameter values and their initial values
so that the daily case number
and daily death number are best-fitted to the data. For this
reason we will refer to our model as the open SICM model or SICM
model for short.

The last assumption that is not typically made for extended SIR
models is about the daily test-positive rate $P(t)=pCI/(I+aM)$. The
justification is based on Holling's disk function on predation from
theoretical ecology (\cite{Holl1959,Murd1973,Lawt1974,Loga2009}).
Holling's theory, derived for
predation, is universal to all processes involving two entities
one of which must take time to change the encountering of both
into something else. In our setting, Covid-19 testing is an agent
or infrastructure which is to find out infected individuals by testing, namely
the confirmed class $C$, and to find out if an infected individual
under monitoring is free of the virus and thus can be released to the
recovered class $R$. For the first class, there is a discovery
probability rate $a_1$ of the infected class $I$ that will be tested
and confirmed. For the second class, there is a repeating test
rate $a_2$ which is the average number of test an individual
will receive over an average period of days under monitoring. For both cases,
there is an average time $h$ needed to complete a test. Under these
assumptions, the number of daily cases confirmed is the following
Holling Type II function
\[
\hbox{$\frac{a_1I}{1+a_1 h I+a_2 h M}C \quad \Rightarrow
\quad \frac{(1/h) I}{ I+(a_2/a_1) M}C=\frac{pC}{I+aM}I=P$}
\]
where $p=1/h$ is the rate of testing and $h$ is testing time,
$a=a_2/a_1$, the ratio of testing rate
for monitored and infected. The number 1 in the denominator is dropped
for the reason that the numbers of $I$ and $M$ in the denominator
are several orders of magnitude greater. Alternatively, one can start
with the assumption that the daily confirmed number is proportional to
the product of the infected and the confirmed because one class
has a positive feedback on the other class. Because testing takes time,
therefore, the daily rate must be constrained by the time allowed
and the constraining factor is exactly in the form of the denominator
by Holling's theory. For example, on a day no testing is allocated for
the monitored ($a_2=0$), and all testing is done for the infected $I$ which
is very large. Then as $I\to\infty$, the maximum number of test which the
testing infrastructure is capable of completing in a day is $1/h=p$,
the daily testing rate unitized by each confirmed case, with
the maximal confirmed $pC$ per day.
It is the saturation testing rate, implying that when
$I$ is large, not every infected can be tested in a fixed
time. Letter $p$ for the rate connotes PCR test for the virus.

\begin{figure}[t]
\vskip -.5in

\centerline
{\scalebox{.7}{\includegraphics{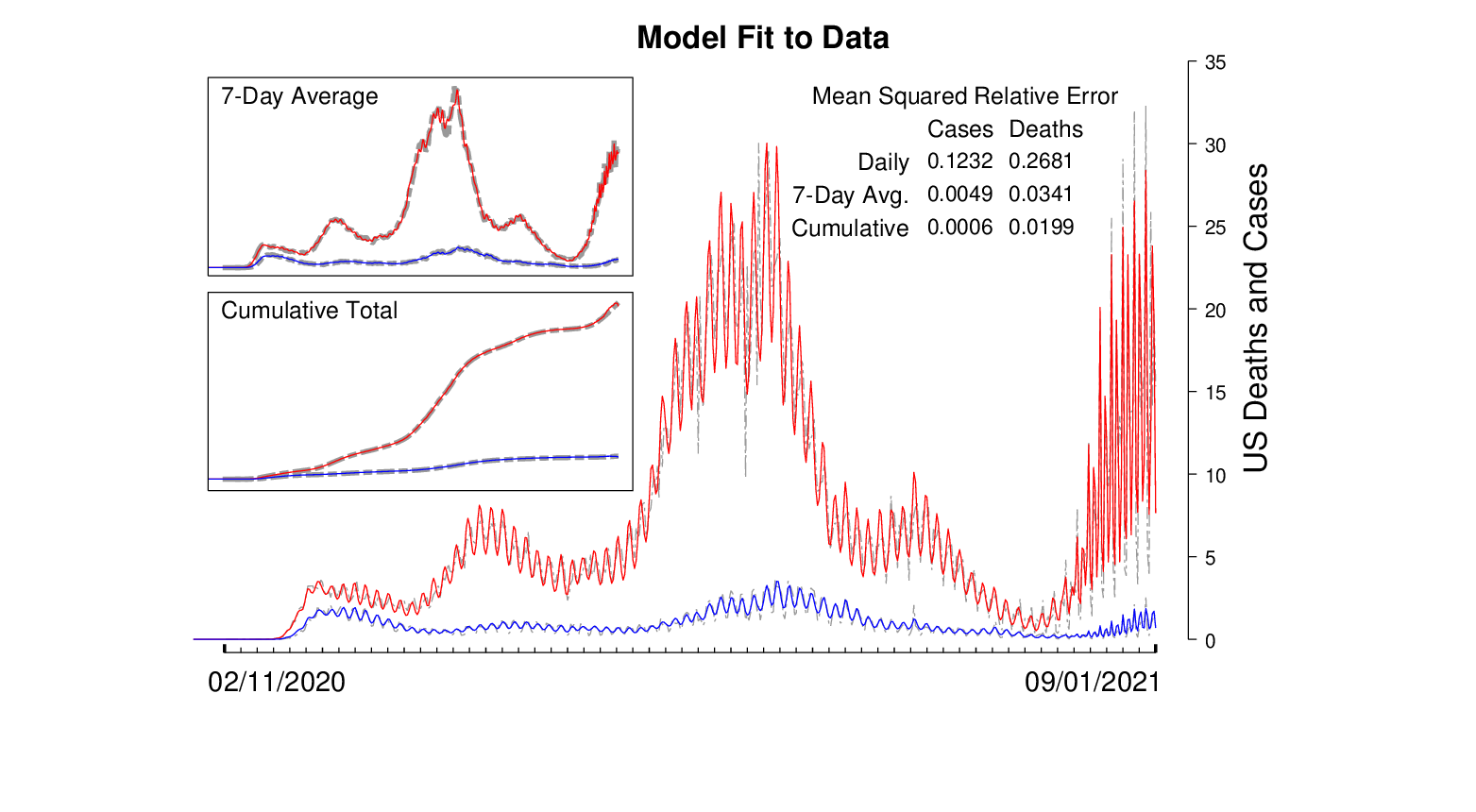}}}
\vskip -.5in
\caption{\textbf{Data-Fitting:} The unit for the fitted
case number (red)
is 1 for $10^4$, and the unit for the fitted death number (blue)
is 1 for $1.25\times 10^3$. For example, the number
4 tick-mark on the scale represents 40,000 for cases
and 5,000 for deaths. This false-scale for plot is used
to boost the visibility of the death data. All plots
use the same case-to-death plot ratio. All dot-dashed
curves (gray) are real data.} \label{figBestFit}
\end{figure}

\bigskip\noindent
\textbf{Data-Fitting.} On each working $k$th day, two tasks are performed
in sequence. The first task is to solve a nonlinear regression problem
to find parameter values and initial conditions
so that the model (\ref{SICM}) is
best-fitted to the data over the past 3 weeks: $(k-\matcdays)\le t\le k$,
with $\matcdays=21$ days.
By best-fit it is meant in theory to find the global minimum
of an error function. But in practice, there is no practical
algorithm to do so. Instead, by best-fit we mean
many local minimums with their fitting or regression errors
ranked from low to high with the lower the error the better the fit.
On each working day, a large number of best-fit searching
is carried out and the best 30 results are ranked and archived.
All results described below are based on the first 5 best-fits
for the 30 saved for each working day, and 541 days
in total from Mar. 12, 2020 to Sept. 1, 2021. The reason that
the first working day starts on Mar. 12 is because
we need to give 21 past days to the first working day
to find its best-fits. We could start the first working
day earlier but the data from the first couple of weeks
were too scattered to be useful.

The searched best-fits are visually represented on the $(-)$-day
columns left of the 0-day column on the diary sheet and
on the record sheet in Fig.\ref{figForecastBook}.
Every variable, every parameter, every quantity and number
we keep track of such as the daily case number,
the daily death number, their seven-day averages
(SDAs), cumulative totals,
etc., each has one diary sheet as
a page to a book. Each book corresponds to one
best-fit searching result. Thus, each working day
generates a volume of 30 books of diary sheets,
and by the same process each working day generates another
volume of 30 books of record sheets. All analyses
and plots are generated from the record sheet
books of the first 5 volumes.

Fig.\ref{figBestFit} shows the result of how our SICM model is fitted to
the U.S. case and death data. Here is how the graph is assembled. All
are based on the daily numbers for cases and deaths. Take the main
daily-case fitting curve for example. For each point
$k$ on the horizontal axis corresponding to the $k$th-day,
we first find the average of the entries from the $k$th-day
row and left of the 0-day column on the daily-case record sheet
for each best-fit volume $1\le i\le 5$, and then average the
5 averages to obtain the plot value for the fitted daily-case
number. Connect these dots from day 50 to day 590 to obtain the
spiking daily-case curve. Once the daily-case sequence is
obtained, the SDA is generated by averaging the past 7 days'
daily-case numbers for each $k$th-day. Similarly, the cumulative
total is generated by summing all past days' daily-case numbers
to day-$k$ for each $k$th-day. The same operations are
applied to obtain these types of curves for deaths.
These definitions for daily number time sequences, SDA number
time sequences, and cumulative total time sequences
will be used throughout for both fit and for forecast
analyses and plots.

The definition for the mean squared relative error (MSRE)
is defined as follows. Let $\delta=\{\delta_1,\delta_2,
\dots,\delta_n\}$ denote any data sequence, such as the daily-case
data, the SDA data, or the cumulative total data, etc.
Let $\phi=\{\phi_1,\phi_2, \dots,\phi_n\}$ denote any fit
or forecast sequence of the same type. Then the MSRE between
the fit and the data is defined as
\begin{equation}\label{defMSRE}
\hbox{$E(\phi,\delta)={\rm mean}\{|\phi_i-\delta_i|^2/\delta_i^2:1\le i\le n\}
=\frac{1}{n}\sum_{i=i}^n{|\phi_i-\delta_i|^2}/{\delta_i^2}$.}
\end{equation}
Fig.\ref{figBestFit} shows, for example, the MSRE for the daily-case
is 12.32\%, for the SDA is 0.5\%, and for the total is
0.06\%, over the time period from day 50 to day 590.
MSRE measures the squared relative error per datum on average.

The method for best-fit is based on Newton's gradient
search (Methods and \cite{bdeng2018,bdeng2019,Rusz2006}).
It is to find the parameter and initial values
of the SICM model so that the MSRE for the joint
daily case and daily death between model's solution and
the data (over the fitting period from $k-\matcdays$
to $k-0$ ) is a minimum. As mentioned before, only
30 best-fits are ranked and archived.

\begin{figure}[t]
\vskip -.5in

\centerline
{\scalebox{.7}{\includegraphics{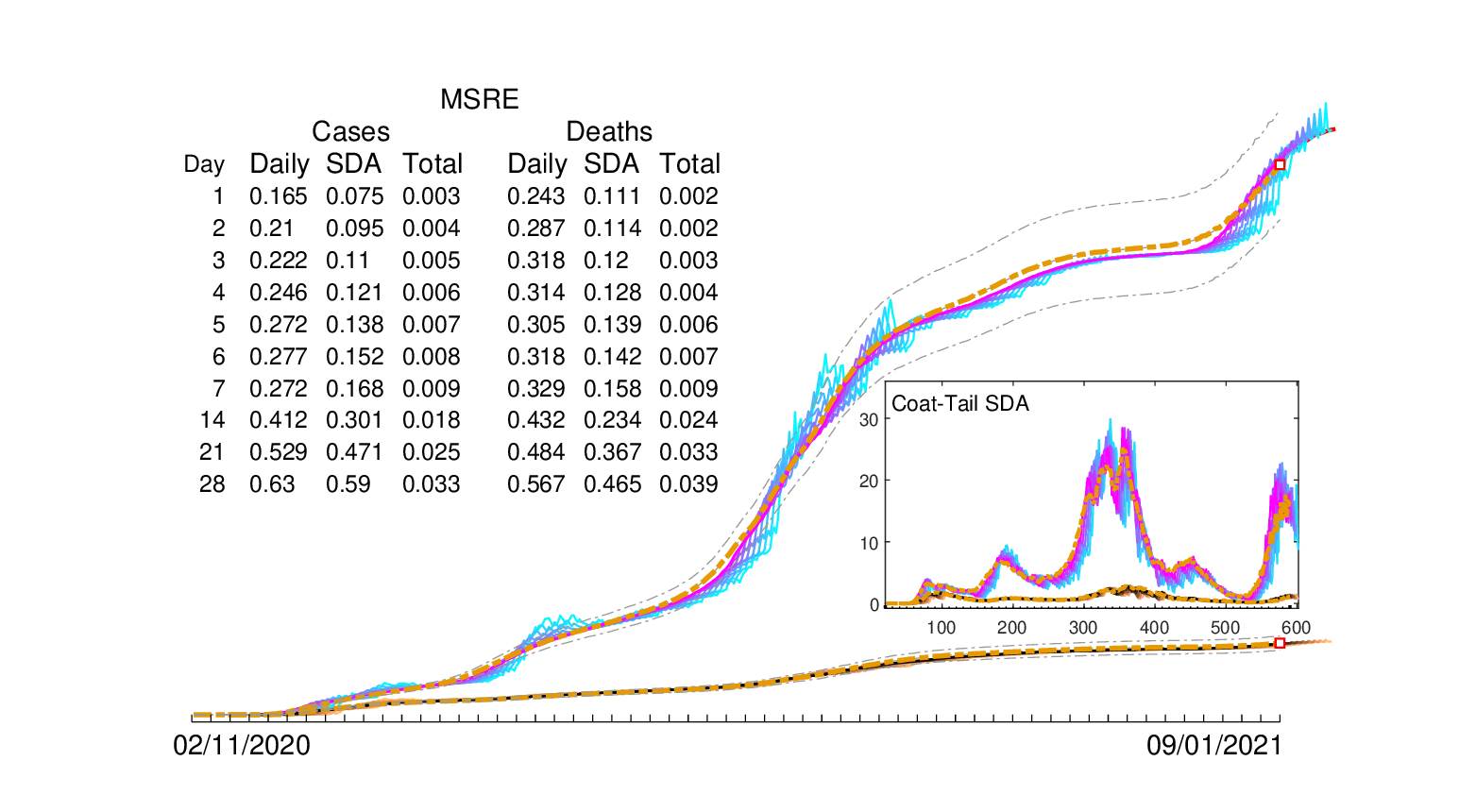}}}
\vskip -.2in
\caption{\textbf{Coat-Tail Plot:} Dashed lines are for real data.
Thin dash-dot lines outline the $\pm$10\% bounds
off the real data. The same conventions apply to
all plots. All `Mean' errors for the paper are MSRE.
The main graph for cumulative totals shows
the primary projections four-weeks ahead
each working day. But the insert for the SDA
coat-tail projection shows only two-weeks ahead
each working day. Not all 28 projected curves are plotted
for a better visibility. For both total and
SDA, the first curve is for the $(+3)$-day projection,
and then every 3rd day projection thereafter. For
the main plot, the last curve ends wit the $(+27)$th-day's
projection and for the insert it ends with the $(+15)$th-day's
SDA.
} \label{figCoatTail}
\end{figure}

\bigskip\noindent
\textbf{Coat-Tail Plot -- Forecaster’s Trap.} On any given
$k$th working day, the second
task following best-fitting is to make a forecast on daily
case and death numbers $\foredays=28$ days into the future.
By definition, all preparatory analyses are referred to as
projections, and only the final projection is called
the forecast. All projections are based on the
columns right of the 0-column on the record sheets
(c.f. Fig.\ref{figForecastBook}).

The first type projection is called the primary projection
or Type-I projection.
It is simply the continuation of solution of
the best-fitted SICM model 28 days ahead for each
$k$th forecasting day. All types of numbers are functions of
of the daily numbers.

Take the daily cases for example. Instead of stopping at
the working day $k$, we define
\begin{equation}\label{defchi}
\chi_{cs,k,i}(+j)=P_i(k+j),\quad  1\le j\le \foredays
\end{equation}
to be the Type-I projection,
where $P_i(t)$ is the model's daily cases projected into
the future on the $k$th forecasting day for the $i$th
best-fitted search, and the first subscript $cs$ is for
`cases'. By extension,
we will use $\chi_{ds,k,i}(+j)=d_iC_i(k+j)$ for the daily
deaths with the subscript $ds$ for `deaths',
and later $\chi_{x,k,i}(+j)=x_i(k+j)$ for any
of the variables $x=S,I,C,M$, etc. The
plus sign `$+$' is used to emphasize the fact
that the quantities are projected future values.

Assume the primary projection is forecasted. The question
for everyone, the forecaster and all observers, is how good
is the forecast? To this question there lies a forecaster's
trap epitomized by what is referred to as a coat-tail plot as
shown in Fig.\ref{figCoatTail}. Take the cumulative total
plot for cases for example. On a forecasting day $k$, the forecaster
uses the known $k$th-day's data $\delta_{t,cs,k}$, with
the first subscript $t$ for `cumulative total', and then add up
the projected daily numbers to obtain the
total for each best-fit $i$:
\[
\hbox{$\psi_{t,cs,k,i}(+z)=\delta_{t,cs,k}+\sum_{j=1}^z\chi_{cs,k,i}(j)$}
\]
for $1\le z\le \foredays$ and $1\le i\le 30$.
Average over $1\le i\le 5$ to obtain the forecasted cumulative total
\[
\psi_{t,cs,k}(+z)={\rm mean}\{\psi_{t,cs,k,i}(+z): 1\le i\le 5\}.
\]
There are 28 projected curves for the cumulative total. For
the $(+1)$-day projection curve, the plotted point for the $k$th-day
is $\psi_{t,cs,k-1}(+1)$ because it is projected on the $(k-1)$th
working day. In general, For the $(+z)$-day projection curve for
$1\le z\le 28$, the plotted point for the $k$th-day
is $\psi_{t,cs,(k-z)}(+z)$ because it is projected on the $(k-z)$th
working day. As for the MSRE, it is calculated as
$E(\psi_{t,cs,\cdot-z}(+z),\delta_{t,cs,\cdot})$ where
the $\cdot$ denotes the position index for elements of
the sequences, $50\le j\le 590$.
The MSRE does not accumulate because the projected sequence and
the data sequence are the same for $j<(k-z)$.

\begin{figure}[t]
\vskip -.5in

\centerline
{\scalebox{.7}{\includegraphics{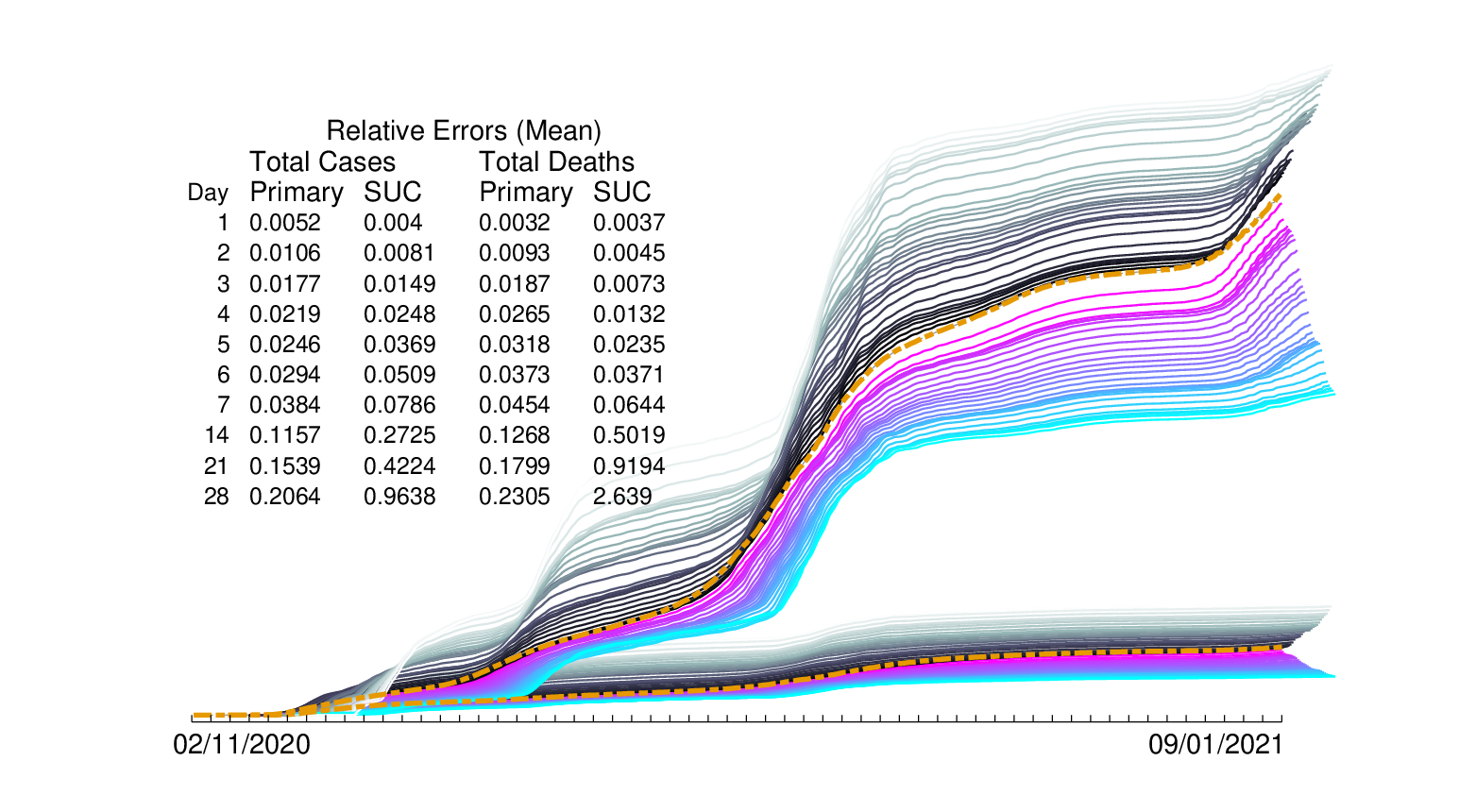}}}
\vskip -.2in
\caption{\textbf{Observer's Plot and Abberations:}
Colored curves are from the primary (Type-I) projections,
with curves closer to the real data being projections closer to
the 0-day. Graded-gray curves (Type-II projection)
are generated by the Step-Up
Correction to the primary projection, with darker the
shade the closer to the 0-day. The primary projection
is an underestimate
abberation and the SUC is an overestimate abberation.} \label{figObserver}
\end{figure}

For the SDA plot in Fig.\ref{figCoatTail}, similar
coat-tail construction applies. Take the case number
for example. We start by considering the solution
of the best-fitted model over the time interval
spanning both the matching window
and the projecting window, followed by computing the SDA
\[
\psi_{s,cs,k,i}(+z)={\rm mean}\{P_i(k+j):
z-6\le j\le z\},\ \hbox{ for $1\le z\le\foredays$}
\]
where $P_i$ is the same daily cases function for the $k$th-day's
best-fit number $i$, $1\le i\le 30$. For the plot, the $k$th-day's
SDA point is the averaged value over the first 5 best-fits
with projection made on the $(k-z)$th working day:
\[
\psi_{s,cs,(k-z)}(+z)={\rm mean}\{\psi_{s,cs,(k-z),i}(+z): 1\le i\le 5\}.
\]
Once the SDA curve is obtained, the mean error
can be computed against the SDA real data
$\{\delta_{s,cs,j}:50\le j\le 590\}$ as we do for the
cumulative total.

The coat-tail forecast is a subjective, forecaster-centric way
to promote the forecaster's methodology. It is prevalent in the
literature (e.g. \cite{bdeng2021,Petr2022}). It is easy to see
from the illustration Fig.\ref{figCoatTail} that a
coat-tail plot can be generated
on any day without reaching deep into the past.
Its projection is attached to the forecaster's
model-fitting black box -- all columns left of the
0-day column, and hence the `coat-tail' moniker. The
coat-tail projections on the SDA and cumulative total
do not accumulate in errors, which always result
in a bias forecasters often fail to recognize.

\bigskip\noindent
\textbf{Observer's Plot and Abberations.}
To evaluate the performance of
a projection, there is another way other than the
coat-tail plot preferred by the forecaster. It is the
observer's plot. To evaluate how good a $(+7)$-day
forecast is, for example, an observer only needs to compare
all $(+7)$-day forecasts by the forecaster to
the actual data. Take the cumulative total for cases
as an example again. This is to
compare the actual data sequence
$\delta_{t,cs,k}$ with the projected data
$\phi_{t,cs,(k-z)}(+z)$ with $z=+7$ for all past forecasting days where
\[
\phi_{t,cs,(k-z)}(+z)=\mean\{\phi_{t,cs,(k-z),i}(+z):1\le i\le 5\}
\]
and
\begin{equation}\label{defphi}
\hbox{$\phi_{t,cs,(k-z),i}(+z)=\sum_{j=z+1}^{k+z}\chi_{cs,(j-z),i}(+z)$}
\end{equation}
where $\chi_{cs,j,i}(+t)$ is the working $j$th-day's
projected daily cases from definition (\ref{defchi}).
The curve $\{\phi_{t,cs,(k-z)}(+z):(z+1)\le k\le 590\}$
with $z=+7$ by definition is the observer's plot for
forecaster's $(+7)$-day performance, and the MSRE
is computed similarly by
\[
\hbox{$E_{t,cs}(+z)={\rm mean}\{(\phi_{t,cs,(j-z)}(+z)
-\delta_{t,cs,j})^2/\delta_{t,cs,j}^2: (z+1)\le j\le 590\}$}
\]
for $z=+7$. The observer's evaluation on any other
$(+z)$-day's forecast for $1\le z\le \foredays$ is done
similarly. In particular, the SDA for cases, for example, is given by
\[
\hbox{$\phi_{s,cs,k,i}(+z)=\mean\{\chi_{cs,j,i}(+z): k-6\le j\le k\},\ k\ge 7$}
\]
for each best-fit $i\le i\le 30$. For each $1\le z\le \foredays$
the $\phi$ curves as time series in $k$ are
referred to as the observe's curves for the $(+z)$-day.

Fig.\ref{figObserver} contains
all 28 primary projections for case total and 28
primary projections for death total, with e.g.
$E_{t,cs}(+7)=0.0384$.
For the $(+28)$-day projection, the MSRE is
$E_{t,cs}(+28)=0.2084$, but from the plot we can see
that for many parts of the curve, the error is way
greater than 20\%, pushing to 40\% and 50\%. In comparison
with its coat-tail counterpart from Fig.\ref{figCoatTail},
the $(+28)$-day's projection error is only 5\% on average
and rarely exceeds the $\pm 10$\% envelope.
To summarize, an observer's curve must be generated
deep into the past, without connection to
forecaster's `black box', and without constant
resetting by the forecaster to the real data like
its coat-tail counterpart does, its error from the real data
almost always accumulates.

But the type of error we see from observer's plots is
not random. Instead, it is a rather simple abberation
because the $(+)$-day projection for the
cumulative totals is a consistent deviation from a norm,
namely the data. Specifically,
each curve is an under-estimate of the data, apparent
from the very beginning, an insight extremely useful
for the forecasters, and the further ahead of the
projection the greater the deviation for the projection
from the data. Like all abberations, it can be
corrected to various degrees.

\begin{figure}[t]
\vskip -.5in

\centerline
{\scalebox{.7}{\includegraphics{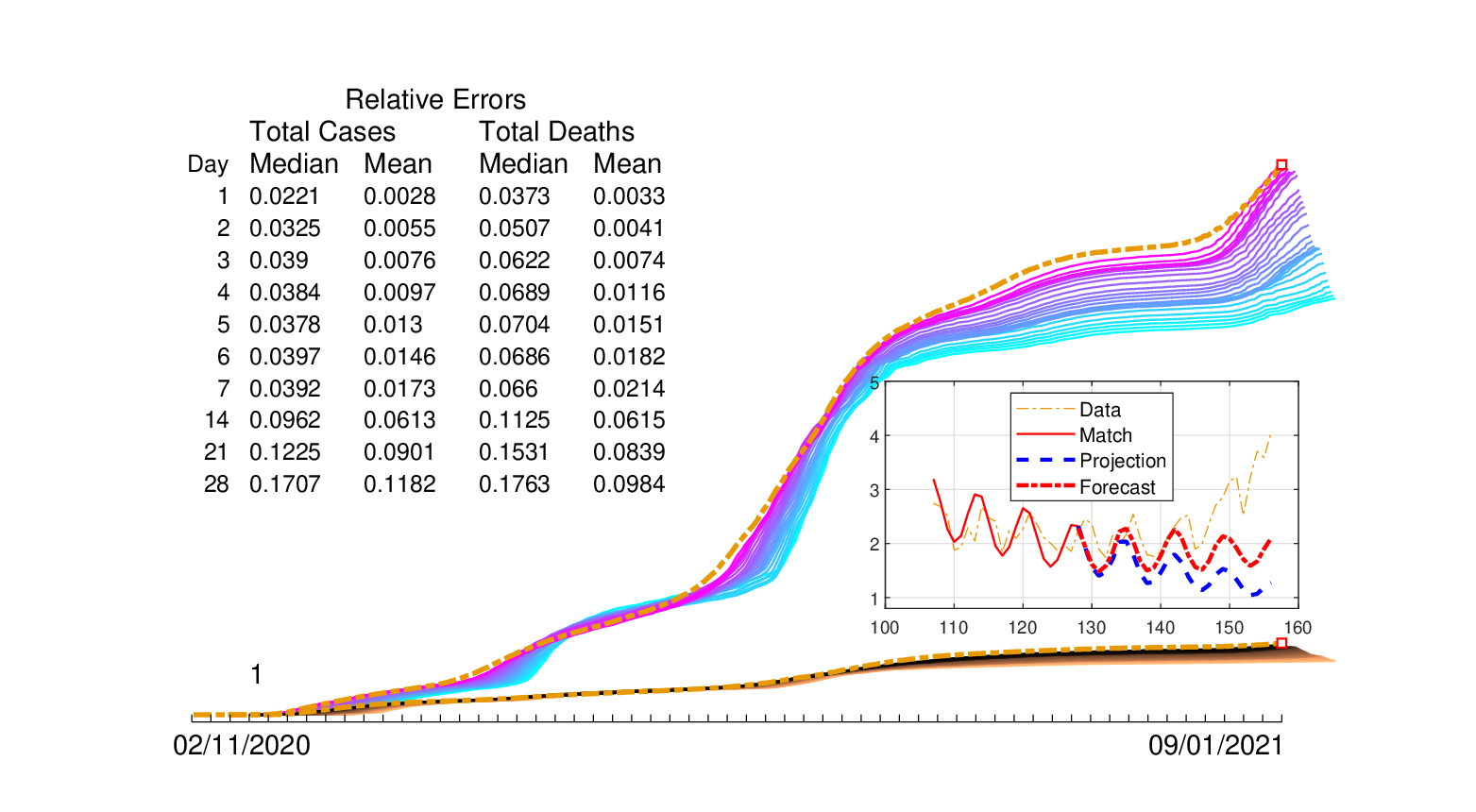}}}
\vskip -.2in
\caption{\textbf{Abberation-Corrected Projection:}
The AAC projection (Type-III) for most of the
working days is an under-abberation.
The $b$ value is $b=1$, at the bottom left corner.
The median error is defined as the median of
the relative errors:
${\rm median}\{|\phi^1_{t,y,j-z}(+z)
-\delta_{t,y,j}|/\delta_{t,y,j}: 50\le j\le 590\}$
for $y=cs,\ ds$. Insert is for
a sample result for the case numbers from the 128th working day:
showing the model-fitted curve to data (solid red),
the Type-I projection (dashed blue), and
the Type-III forecast (dot-dashed red).}
\label{figAACbeta1}
\end{figure}

\bigskip\noindent
\textbf{Boosted Abberation Corrections for Forecast.}
The second type projection (Type-II) is called the
step-up-correction (SUC) projection. It is based on
the cumulative totals for the data $\delta_{t,y,k}$ and for the
primary projection $\phi_{t,y,(k-z)}(+z)$, $y=cs, ds$.
We will use the superscript `$suc$'
to denote such projections.
Take the cumulative total for the cases for example. Because the
primary projection for $(+z)$-day,
$\phi_{t,cs,(k-z),i}(+z)$
is an underestimate of the real data,
$\delta_{t,cs,k}$, c.f. Fig.\ref{figObserver},
the idea is to correct the abberation by lifting up
the projection. As always, we start with the daily
case number, $\chi_{cs,k,i}(+z)$, which is scaled up
by a compensating factor which is the ratio of data
to projection in the case totals:
\[
\hbox{$\chi^{suc}_{cs,k,i}(+z)=\frac{\delta_{t,cs,k}}
{\phi_{t,cs,(k-z)}(+z)}\chi_{cs,k,i}(+z)$}
\]
on the $k$th forecasting day for all $1\le z\le\foredays$
with $\chi$'s definition from (\ref{defchi}). Similarly,
we will extend the definition to all variables $x=S,I,C,M$
$\chi^{suc}_{x,k,i}(+z)=[\delta_{t,cs,k}/\phi_{t,cs,(k-z)}(+z)]
\chi_{x,k,i}(+z)$, except for the daily death number
which is defined as
\[
\hbox{$\chi^{suc}_{ds,k,i}(+z)=\frac{\delta_{t,ds,k}}
{\phi_{t,ds,(k-z)}(+z)}\chi_{ds,k,i}(+z)$}
\]
We can then define the SUC for
$\phi^{suc}_{t,cs,k,i}(+z)$ the same as we do in (\ref{defphi})
and
\[
\phi^{suc}_{t,cs,(k-z)}(+z)=\mean\{\phi^{suc}_{t,cs,(k-z),i}(+z):1\le i\le 5\}
\]
and the MSRE $E^{suc}_{t,cs}(+z)$ similarly as for
$E_{t,cs}(+z)$ as above. We can also define similarly
for SDA related quantities: $\phi^{suc}_{s,cs,(k-z)}(+z)$,
$E^{suc}_{s,cs}(+z)$, etc. And similarly for the
death numbers and MSREs by substituting the subscript
$cs$ by $ds$. Fig.\ref{figObserver} also
shows the $\phi^{suc}$ curves for the cumulative totals
for cases and deaths. It also shows that the MSRE $E^{suc}$
is worse off for $(+)$-days further away from the
0-day. Fortunately, the SUC projection is again an
abberation. But this time it is an
overestimate abberation, which again
is subject to corrective intervention.

\begin{figure}[t]
\vskip -.5in

\centerline
{\scalebox{.7}{\includegraphics{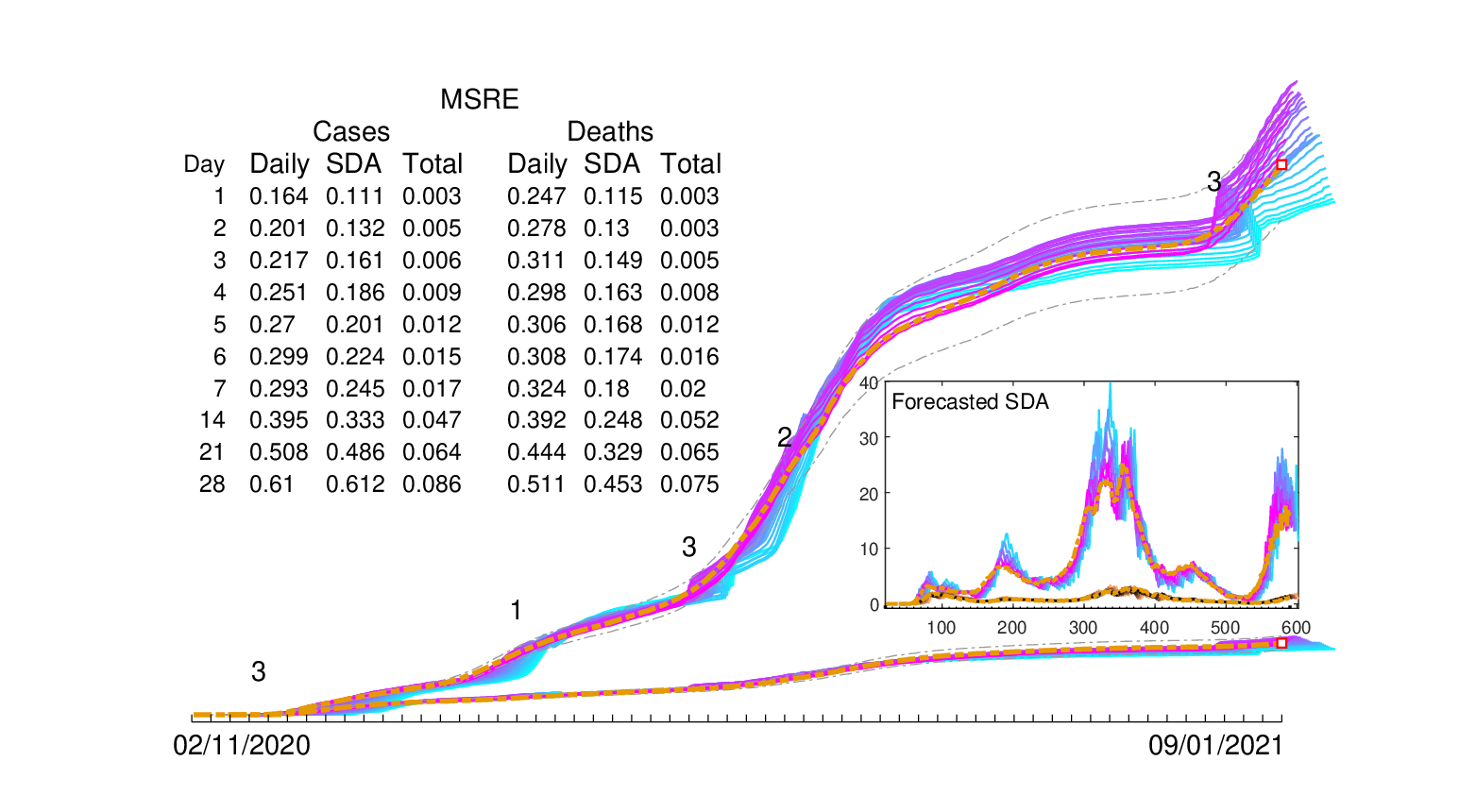}}}
\vskip -.2in
\caption{\textbf{Forecasting:} Use different boosted AACs
to make the daily forecast. From day 50 to day 185, $b=3$
is used, from day 186 to 275, $b=1$ is used, and
the subsequent switching takes place on day 325, 550 with
the $b$ value equal 2, 3, respectively.
The main plot is the cumulative totals for all 28 days.
Insert shows the corresponding SDA forecast, plotted
every plus-three days for 15 days ahead on each
working day. Daily forecast is
not shown but their daily variations are
comparable with their SDA counterparts as shown
in the MSRE table.} \label{figBoost}
\end{figure}

The Type-III projection is referred to as
averaged-abberation-correction (AAC) or boosted AAC,
and we will use a superscript value `$b$' to denote such
projections. Again, take the cumulative total
for cases as example and start with the daily
case number. The AAC projection for
the daily number on the $k$th-forecasting
day is to apply a weighted-average to
the primary under-abberation
projection and the SUC over-abberation projection
as follows
\begin{equation}\label{defAAC_1}
\hbox{$\chi^{1}_{cs,k,i}(+z)=\frac{\weightp\chi^{suc}_{cs,k,i}(+z)
+\weightsuc\chi_{cs,k,i}(+z)}{\weightp+\weightsuc}$ where}
\end{equation}
\begin{equation}\label{defweight}
\hbox{$\weightp=\phi_{t,cs,(k-z)}(+z)$ and $\weightsuc=
\phi^{suc}_{t,cs,(k-z)}(+z)$.}
\end{equation}
That is, the daily
over-abberation is weighted by the primary
under-abberation total and the daily under-abberation
is weighted by the SUC over-abberation total.
Here $b=1$ for this AAC.
We can then use the AAC daily case numbers to
define $\phi^{b}_{t,cs,(k-z)}(+z)$ and
$E^{b}_{s,cs}(+z)$ for $b=1$, etc.
As shown in Fig.\ref{figAACbeta1}, the AAC
is a substantial improvement over the Type-I
and the Type-II projections. For example,
the (+28)-day projection's MSRE for
the cumulative case total is only 11.8\% and that
for cumulative death total is below 10.0\%.

Because this AAC projection for the case total with $b=1$ is still
an under-estimate abberation, the SUC projection for the case total
can afford to be an even greater over-abberation. We can
do this by scaling up the weight $\weightsuc=
\phi^{suc}_{t,cs,(k-z)}(+z)$ in (\ref{defAAC_1})
by a boosting factor.
Specifically, let $b$ be $b\ge 1$ and
for any $(+)$-day number $+z$, $1\le z\le\foredays$,
define the incremental boosting factor as
\[
\beta(+z)=1+\Delta_b(z-1),\ \hbox{ where $\Delta_b=\frac{b-1}{\foredays-1}$}
\]
so that $\beta(+1)=1$ is for no boost, $\beta(+2)=1+\Delta_b$ is for
one $\Delta_b$ boost, etc, and $\beta(+\foredays)=b$ is for the full
$b$ amount boost as $1\le \beta(+z)\le b$. The daily case
number for the boosted AAC projection with $b\ge 1$ is
\begin{equation}\label{defAAC_12}
\hbox{$\chi^{b}_{cs,k,i}(+z)=\frac{\weightp\chi^{suc}_{cs,k,i}(+z)
+\beta(+z)\weightsuc\chi_{cs,k,i}(+z)}{\weightp+\beta(+z)\weightsuc}$}
\end{equation}
which generalizes the definition (\ref{defAAC_1}) to
any value $b\ge 1$. From these boosted daily case numbers
we can define the corresponding boosted AAC projection
for the case total $\phi^{b}_{t,cs,(k-z)}(+z)$ and its MSRE
$E^{b}_{s,cs}(+z)$ for $b\ge 1$. The same can be done for
the boosted AAC projections on the daily death number
\[
\hbox{$\chi^{b}_{ds,k,i}(+z)=\frac{\weightp\chi^{suc}_{ds,k,i}(+z)
+\beta(+z)\weightsuc\chi_{ds,k,i}(+z)}{\weightp+\beta(+z)\weightsuc}$}
\]
using exactly the same weights, and subsequently the death total,
$\phi^{b}_{t,ds,(k-z)}(+z)$,
the SDA numbers,  $\phi^{b}_{s,y,(k-z)}(+z)$,
and their MSREs $E^{b}_{x,y}(+z)$ with $x=d,s,t$ and $y=cs,ds$.

At last on any $k$th forecasting day, here is how the
forecast is generated of which the result is shown in Fig.\ref{figBoost}.
After the $k$th-day's data becomes available, search for
best-fits is carried out
and the best 30 are ranked and archived. Then the Type-I
projection (primary) is made, followed by the Type-II
projection (SUC). Then for a few booster values, for example,
$b=1,2,3,4,\dots$ or fractions of integers, $1,1.5,2$, etc.,
the Type-III projections are made. We then
exam the performance of these boosted AACs for the past two
weeks, and choose one boosted Type-III projection for the
day's forecast because of its overall fit to data and
its overall trending. Fig.\ref{figBoost} shows that
we only need to switch the boosting
value 5 times for the simulated period. It shows that
we can keep the MSREs for both case and death totals  within 10\%
of the real data, comparable to the biased coat-tail result.

\bigskip\noindent
\textbf{Methods.}
The Matlab data that support the findings of this study are available in figshare
with the identifier doi.org/10.6084/m9.figshare.21968660.v2 (\cite{bdeng2023}).
It requires zero knowledge to verify the results represented by all figures of this paper.
All one needs is the model file `SICM.m' for Eq.(\ref{SICM}), the initial variable
and parameter values
from the data file `Input\_Matched\_InitialsAndParameters.mat'  to explore the dynamics
of the model, and to analyze its fit to the CDC data  (\cite{CDC21}) which is
included to the data set. In addition, for the 3-week daily-fit as illustrated in the insert
of Fig.\ref{figAACbeta1} a 590-day animation is given by Daily\_Match\_Data\_to\_Model.mov.
For the best-fit from Fig.\ref{figBestFit} an animation is given by Match\_Data\_to\_Model.mov,
and for the simulated forecast Fig.\ref{figBoost} an animation is given by SICM\_Forecast.mov.

The initial variable and parameter values for the best-fit of the model to the data are
found by a global optimization algorithm implemented in Matlab. Such an algorithm can be
any program or function mfile contained in the Matlab library, such as the package of
GlobalSearch and MultiStart, Genetic Algorithm, Particle Swarm,
Multiobjective Optimization. All are variants based on Newton's
gradient search idea. Ours can be categorized as a global line-search algorithm using
the ideas of genetic algorithm, multi-start, particle swarm, and multi-objective optimizations.
The search is done against the loss function
\[
L(ip)=E(\chi_{d,cs},\delta_{d,cs})+wE(\chi_{d,ds},\delta_{d,ds})
\]
where $ip$ denotes the initial and parameter values of the SICM model, $E$ is the MSRE defined in
(\ref{defMSRE}), and $w>0$ is a weight parameter to adjust the amounts of the case's MSRE and the
death's MSRE in the loss function. The first subscript `$d$' above is for
the daily numbers of both cases ($cs$) and deaths ($ds$). The idea of line-search is explained in
\cite{bdeng2018,bdeng2019,Rusz2006}. Instead of the true gradient of the loss function, we follow
the fastest descent along an initial variable's coordinate or a parameter's coordinate
at each iteration in an interval twice the coordinate value with a preset search step-size.
As we mentioned above this is just one out of many methods to leverage the
computational power in speed and memory of computers in order to accomplish the same goals.

\begin{figure}[t]

\centerline{\scalebox{.65}{\includegraphics{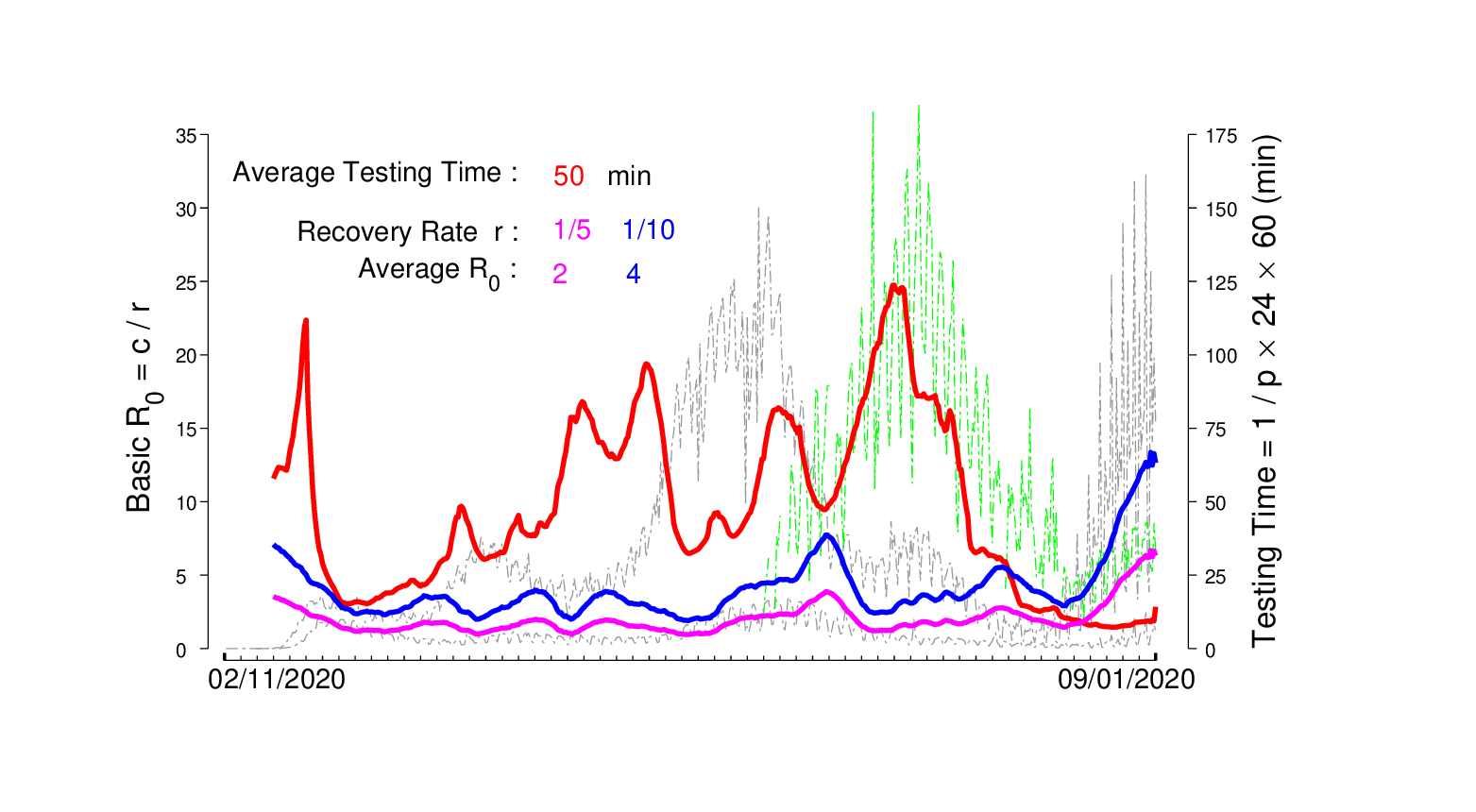}}}
\vskip -.4in

\caption{\textbf{Future Study:} A preview to post-mortem analysis on the
pandemic. Notice that the basic reproduction number
climbed up with the appearance of the omicron variant in
the months of July and August, 2021. Background data in green
is the daily vaccination numbers in unit of
$1.25\times 10^5$ in the same plot scale as the data
in Fig.\ref{figBestFit}. The $R_0$ number is consistent
with other's estimates (e.g. \cite{Zou2020}).} \label{figDiscussion}
\end{figure}

\bigskip\noindent
\textbf{Discussion.} We can further consolidate the performance
of our method to one value for each case total and death total:
the median value of their MSRE (Fig.\ref{figBoost}).
That is, we can say the median MSRE
error for a 28-day forecasting is 0.0473 (the (+14)-day's MSRE)
for the case total
and 0.052 for the death total. By exactly the same definition,
the median MSRE error for a 28-day forecasting is
0.395 for the daily cases and 0.3335 for the SDA cases,
and, 0.392 for the daily deaths and 0.2481 for the SDA deaths,
both are expected to have greater variations than their
cumulative totals, which tend to smooth out short-term
volatilities. In comparison, a forecaster's biased
coat-tail median MSREs (Fig.\ref{figCoatTail}) are
smaller than or around half the observer's values.

There are many ways to make use of the 30 best-fits
over the 590-day study period. For example, one can use them
as initial guesses to fit their own models to other data, over
longer period of time, for real time forecasting, for different states or
countries. For another example, one can use them to do post-mortem
analyses on the pandemic. Fig.\ref{figDiscussion} shows a preview
to such studies. It shows if the average recovery time is 10 days,
then on average each infectious person will pass the virus to 4
persons. We intend to publish such a study in the future.

Although the model dynamics of Fig.\ref{figModelDynamics} is not
meant for the real data, the parameter values are taken from
one of the best-fits. It appears to show that the virus can cause
an outbreak for 100--200 days. This information
on its own may not mean much, but when it is viewed together
with the real data over the 590-day study period, it seems to
suggest that on average the outbreak caused by
each variant does last about 100--200 days.
This points to a future modification of our SICM
model to include multi-variant dynamics to the model. The insert of
Fig.\ref{figAACbeta1} is chosen to show such a need.
It suggests that one outbreak is ending and another outbreak
is starting. Our one-variant SICM model performs poorly
during such overlapping periods, in which its projections
are trending in the wrong direction. With a two-variant
model, such transitions are expected to become smoother,
and possibly further reduce the MSREs for the daily and
SDA numbers shown in Fig.\ref{figBoost}.

I did a real-time forecasting for the New York State's case number
from March 20 to April 30, 2020 (\cite{bdeng2021}). I stopped the experiment when
I realized the simple variant of the SIR model I used was not
capable to generate the seven-day oscillation. This study is a
continuation of that experiment, and our SICM model is an improvement
as a result. However, it is meant to be a minimal model capable of
the oscillation, and there are room for improvements. For example,
because its primary projection consistently under estimate
the real data, the model should be further revised to reduce
the amount of abberations.
One possible modification is to try a Holling Type-II form for the
daily infection rate term $cSI$. Another possible improvement is
to include the daily test number into the best-fit that includes
the number of test-negative individuals from the susceptible group $S$.
These modifications may also reduce the MSREs for the daily and
SDA numbers.

There is a possible immediate \textit{ad hoc} addition to our
daily forecasting regime that may improve the daily and SDA
errors. By comparing the coat-tail daily plot from the
insert of Fig.\ref{figCoatTail} and the boosted AAC
forecast from the insert of Fig.\ref{figBoost}, we can
see that the latter removes almost all the under-abberations
for the SDA numbers but add more over-abberations to
the time periods when the cases reach local maximums. Thus,
the additional rule is to turn off boosted AAC whenever the SDA
exhibits a consecutive overestimate of the data for 3 to 7 days,
and then turn on boosted AAC whenever the primary SDA projection
becomes an underestimate for a few days in a row.

We note that the testing time $1/p$
shown in Fig.\ref{figDiscussion} by
definition is the same as the handling time from
mathematical ecology. This time definitely includes
the time from the moment when a sample is collected
to the moment when the result is entered
into the reporting system which generated the data
for this study. But it should not include
the time taken to inform the patient. It may not
include excessive idling time the sample is not
processed, which as an assumption may be subject
to debate. The last comment before closing is
this, the usefulness of the method depends on
the data and the model. Bad data or bad model
will make the whole exercise pointless.

The methodology we adopted is as traditional as
Newtonian mechanics: construct a model for a system,
fit the model to real data, and make prediction for
the future trajectory of the system. There is not a
better example for this Newtonian tradition than
the planetary model for the solar system.
Obviously, our model for the Covid-19 epidemic is
nowhere as good as many models in physics and engineering.
Our method for best-fitting follows the same tradition
by gradient search, another great idea by Newton.
In our case and for all applications of Artificial
Intelligence where gradient search is the key,
we are simply leveraging the power of computers
for an old idea. The similarity between
our method and the Newtonian tradition ends here. Because
of its accuracy, there is few abberation in the prediction
of planetary motion by Newton's gravitational model, and
hence its forecaster's plot and observer's plot are the same.
In contrast, because of our model's incompleteness,
abberations to projection are bound to happen.
This is perhaps the lesson we learn from this paper,
and as a result to
stimulate others to find better models and to improve
the science of epidemiological forecasting so that we can
be better prepared for the next pandemic.

\bigskip\noindent
\textbf{\large Declarations}

\small\noindent
\textbf{Ethical approval:} Not Applied.

\small\noindent
\textbf{Conflict of interests:} None.

\small\noindent
\textbf{Authors' Contributions:} Not Applied.

\small\noindent
\textbf{Funding:} None.

\small\noindent
\textbf{Availability of data and materials:} All best-fitted data for 
this article can be downloaded from \cite{bdeng2023}. Additional 
materials can be made available from the author on reasonable request.


\end{document}